\documentclass[aps,prl,reprint,twocolumn,amsmath,amssymb,superscriptaddress,showpacs]{revtex4-1}
\usepackage{graphicx}
\usepackage{dcolumn}
\usepackage{bm}

\newcommand{\bise}{Bi$_2$Se$_3$}
\newcommand{\bite}{Bi$_2$Te$_3$}
\newcommand{\sbte}{Sb$_2$Te$_3$}

\usepackage{amsmath}
\usepackage{amssymb}
\usepackage[usenames]{color}
\usepackage[normalem]{ulem}

\begin{document}

\title{Relativistic k$\cdot$p Hamiltonians for centrosymmetric topological insulators\\
from \textit{ab initio} wave functions}

\author{I. A. Nechaev}

\affiliation{Centro de F\'{i}sica de Materiales CFM - MPC and Centro Mixto CSIC-UPV/EHU, 20018 San Sebasti\'{a}n/Donostia, Spain}
\affiliation{Tomsk State University, 634050 Tomsk, Russia}
\affiliation{Saint Petersburg State University, 198504 Saint Petersburg,  Russia}

\author{E. E. Krasovskii}
\affiliation{Departamento de F\'{i}sica de Materiales UPV/EHU, Facultad de Ciencias Qu\'{i}micas, UPV/EHU,
Apdo. 1072, 20080 San Sebasti\'{a}n/Donostia, Spain}
\affiliation{Donostia International Physics Center, 20018 San Sebasti\'{a}n/Donostia, Spain}
\affiliation{IKERBASQUE, Basque Foundation for Science, 48013 Bilbao, Spain}

\date{\today}

\begin{abstract}
We present a method to microscopically derive a small-size
\textbf{k$\cdot$p} Hamiltonian in a Hilbert space spanned by physically chosen
\textit{ab initio} spinor wave functions. Without imposing any complementary
symmetry constraints, our formalism equally treats three- and two-dimensional
systems and simultaneously yields the Hamiltonian parameters and the
true $\mathbb{Z}_2$ topological invariant. We consider bulk crystals and thin
films of Bi$_{2}$Se$_{3}$, Bi$_{2}$Te$_{3}$, and Sb$_{2}$Te$_{3}$. It turns out
that the effective continuous \textbf{k$\cdot$p} models with open boundary
conditions often incorrectly predict the topological character of thin films.
\end{abstract}

\pacs{71.15.-m, 71.18.+y, 71.70.Ej, 73.22.-f}
%
%

\maketitle

Electronic structure of topological insulators (TIs) has been in focus of theoretical
research regarding linear response, transport properties, Hall conductance, and motion
of Dirac fermions in external fields \cite{Qi_RMP_2011, Weng_AIP_2015}. These problems
call for a physically justified model Hamiltonian of small dimension. As in semiconductors,
it is thought sufficient that the model accurately reproduces the TI
band structure near the inverted band gap~\cite{Zhang_Nat_Phys_2009}. The desired
Hamiltonian is derived either from the theory of invariants \cite{Winkler} or within
the \textbf{k$\cdot$p} perturbation theory using the symmetry properties of the basis
states \cite{KP_book}.

In Ref.~\cite{Zhang_Nat_Phys_2009}, along with the pioneering prediction of
the topological nature of \bise, \bite, and \sbte, a 4-band Hamiltonian was
first constructed from the theory of invariants, which is presently widely
used to analyze the properties of bulk TIs as well as their surfaces and thin
films \cite{Linder_PRB_2009, Fu_PRL_2010, Apalkov_PRL_2011, Silvestrov_PRB_2012,
Ebihara_PhysE_2012, Zhu_PRL_2014, Saha_PRL_2015, Orlita_PRL_2015, Liu_PRB_2015}.
The Hamiltonian parameters in Ref.~\cite{Zhang_Nat_Phys_2009} were obtained by
fitting \textit{ab initio} band dispersion curves. Later, an attempt was made
\cite{Liu_PRB_2010} to recover the Hamiltonian of
Ref.~\cite{Zhang_Nat_Phys_2009} by a \textbf{k$\cdot$p} perturbation
theory with symmetry arguments and to derive its parameters from the
\textit{ab initio} wave functions of the \textit{bulk}
crystals. Furthermore, in Ref.~\cite{Liu_PRB_2010} the effective
Land\'{e} $g$-factors for the Zeeman splitting \cite{Winkler, KP_book}
were introduced within the \textbf{k$\cdot$p} theory of TIs.

To analyze how the properties of \textit{thin films} are inherited from the bulk TI
features, effective continuous models have been developed: they are based on the
substitution $k_z \to -i{\partial }_z$ (originally introduced for slowly varying
perturbations \cite{Slater_PR_1949}) in the Hamiltonian of Ref.~\cite{Zhang_Nat_Phys_2009}
and on the imposition of the open boundary conditions
\cite{Liu_PRB_2010, Lu_PRB_2010, Shan_NJP_2010, Zhang_PRB_2012}.
These models predict a variety of intriguing phenomena at surfaces, interfaces,
and thin films of TIs \cite{Lu_PRL_2013, Zhang_PRL_2014, Parhizgar_PRB_2015, Zhang_SciRep_2015}.
A fundamental issue here is the topological phase transition between an ordinary 2D insulator
and a quantum spin Hall insulator (QSHI). Apart from the theoretical prediction, the model
parameters are fitted to the measured band dispersion to deduce the topological phase from
the experiment \cite{Zhang_NatPhys_2010, Zhang_PRL_2013_2}.
By analyzing the signs and relative values of the parameters of the empirically obtained
effective model a judgement is made on whether the edge states would exist in a given TI film, the
logic being similar to that of Ref.~\cite{Zhou_PRL_2008}: The valence band should have a positive
and conduction band a negative effective mass.

In order to avoid any ambiguity in deriving the model Hamiltonian and
to treat 3D and 2D systems within the same formalism, one needs an
\textit{ab initio} and internally consistent scheme that realizes the
\textbf{k$\cdot$p} theory with the full inclusion of the microscopic
structure of the system and generates a compact and physically
transparent form of the Hamiltonian of a given size. A few attempts have
been recently undertaken to predictably construct model Hamiltonians
for classical bulk semiconductors
\cite{Junior_PRB_2016} and graphene-based systems \cite{Ray_arxiv_2016}.

Here, we report a method to \textit{microscopically} derive the relativistic Hamiltonian
$H_{\mathrm{\mathbf{kp}}}$ accurate up to the second order in \textbf{k} from the spinor wave functions
obtained with the all-electron full-potential extended linearized augmented plane wave method (ELAPW).
The size of $H_{\mathrm{\mathbf{kp}}}$ is determined by the dimension of the subspace spanned by the physically
chosen basis states. The form of the Hamiltonian is dictated by the
symmetry of the wave functions unitary transformed to diagonalize the $z$-component of the total
angular momentum and by a universal prescription to choose their phases. Here we apply this approach
to centrosymmetric bulk crystals as well as to thin films of Bi$_{2}$Se$_{3}$, Bi$_{2}$Te$_{3}$, and
Sb$_{2}$Te$_{3}$ up to six quintuple layers (QLs). For each film, we calculate the topological
invariant, indicating whether it is a QSHI. We conclude on the validity of the effective
models by comparing the predictions by the \textbf{k$\cdot$p} Hamiltonian with the actual properties
of the film. Furthermore, within our approach, we derive for the first time the \textbf{k$\cdot$p} Zeeman term for the films.

We construct the model Hamiltonian as a second-order \textbf{k$\cdot$p} expansion around
the point ${\mathbf k}=0$. To avoid any ambiguity, we obtain the expansion coefficients
directly from {\it ab initio} eigenfunctions at $\Gamma$. For systems with inversion
symmetry the energy bands $E_n$ are doubly degenerate with two orthogonal wave functions
$\Psi_{n1}$ and $\Psi_{n2}$ that are parity eigenfunctions at the time reversal invariant
momenta (TRIM).
The \textbf{k$\cdot$p} Hamiltonian is represented in the basis of these functions in terms
of the matrix elements $\langle \Psi_{ni}|\bm{\pi}|\Psi_{mj}\rangle$ of the velocity operator
$\bm{\pi}=-i\hbar\mathrm{\bm{\nabla}}+\hbar\left[\bm{\sigma} \times \mathrm{\bm{\nabla}} V\right]/4m_0c^2$
\cite{Krasovskii_PRB_2014}. Here $\bm{\sigma}$ is the vector of Pauli matrices and $V(\mathrm{\mathbf{r}})$
is the crystal potential. In the \textbf{k$\cdot$p} expansion $H_{\mathrm{\mathbf{kp}}}=H^{(0)}+H^{(1)}+H^{(2)}$,
the zero-order term is just the band energy, $H^{(0)}_{nimj}=E_n\delta_{mn}\delta_{ij}$, the linear
term is $H^{(1)}_{nimj}=(\hbar/m_0)\mathrm{\mathbf{k}} \cdot \bm{\pi}_{nimj}$, and the second-order
term is
\[
H^{(2)}_{nimj}=\frac{\hbar^2k^2}{2m_0}\delta_{mn}\delta_{ij}+\frac{\hbar^2}{m^2_0}\sum_{\alpha\beta}k_{\alpha}D_{nimj}^{\alpha\beta}k_{\beta},
\]
where $\alpha,\beta=x,y,z$, and
\[
D_{nimj}^{\alpha\beta}=\frac{1}{2}\sum_{n'i'}\pi^{\alpha}_{nin'i'}\pi^{\beta}_{n'i'mj}\left(\frac{1}{E_n-E_{n'}}+\frac{1}{E_m-E_{n'}}\right),
\]
see Refs.~\cite{Winkler,KP_book}. Here $m$ and $n$
number the degenerate Kramers pairs, and $i$ and $j$ number the members of a pair.
The index $n'$ runs over all the bands
excluding those forming the \textbf{k$\cdot$p} basis (L\"{o}wdin's partitioning).
Thus, when the dimension of $H_{\mathrm{\mathbf{kp}}}$ equals the dimension of the
original full Hamiltonian the second order term $H^{(2)}$ vanishes \cite{Krasovskii_PRB_2001}
(hereafter, we refer to this case as the full-size \textbf{k$\cdot$p} calculation).

\begin{figure}[tbp]
\centering
\includegraphics[angle=0,scale=0.6]{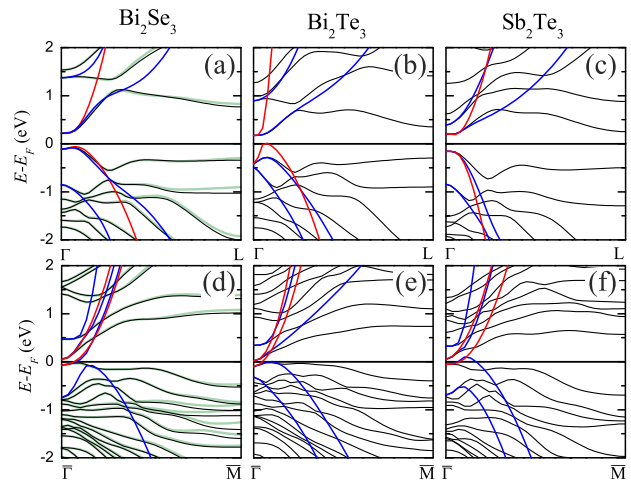}
\caption{Band structure (black lines) of the bulk crystals (a, b, c)
and 2QL films (d, e, f) of \bise, \bite, and \sbte\, compared with the 4-band
(red lines) and 8-band (blue lines) \textbf{k$\cdot$p} model. Results by the
full-size  Hamiltonian are shown by green lines for \bise\,.
See also Figs.~S1--S4 in SM \cite{SM}.}\label{kp_vs_LDA}
\end{figure}

The \textit{ab initio} band structure was obtained with the ELAPW method
\cite{Krasovskii_PRB_1997} using the full potential scheme of Ref.~\cite{Krasovskii_PRB_1999}
within the local density approximation (LDA). The spin-orbit interaction is treated by a
second variation method \cite{KOH77} including the scalar-relativistic bands up to
at least 300 eV. This ensures a good convergence of the inverse effective mass, with a
deviation from the second derivative of the $E({\mathbf k})$ curves within~3\%. The
experimental crystal lattice parameters were taken from Ref.~\cite{Wyckoff} with the LDA
relaxed atomic positions of Refs.~\cite{Nechaev_PRBR_2013, Nechaev_PRB_2013, Nechaev_PRB_2015}.
Figure~\ref{kp_vs_LDA} compares the \textit{ab initio} bands with those obtained by
diagonalizing our \textbf{k$\cdot$p} Hamiltonians of small size (4- and 8-band) and
of full size. Note that the full-size \textbf{k$\cdot$p} calculation highly accurately
reproduces the true bands: the error grows as $k^2$ \cite{Krasovskii_SSC_1995}, and at
the Brillouin zone (BZ) boundary it is within 150~meV.

For each Kramers-degenerate level $n$, the spinor wave functions $\Psi_{ni}$ form
a two-dimensional basis.
Numerically obtained functions are arbitrarily ordered and have unphysical phases,
which, however, affect the structure of $H_{\mathrm{\mathbf{kp}}}$ non-diagonal terms.
In order to keep the same physically motivated ordering and to align the phases in
different calculations we first transfer to the basis that diagonalizes the $z$-component
of the total angular momentum $\mathrm{\mathbf{J}}=\mathrm{\mathbf{L}}+\mathrm{\mathbf{S}}$
in the atomic sphere that has the largest weight in the $n$-th band (see Figs. S5-S7 in the
Supplemental Material (SM) \cite{SM}). This establishes
the numeration of the wave functions $\Psi_{n1(2)}\rightarrow\Psi_{n\uparrow(\downarrow)}$.
Next, we choose the phases of the new basis functions such that they become explicitly
Kramers conjugate: $\Psi_{n\downarrow}=\hat{T}\Psi_{n\uparrow}$, where $\hat{T}=Ki\sigma_y$ is
the time reversal operator and $K$ is the complex conjugation operator.
Finally, for two pairs of different parity, $n$-th and $m$-th, we turn
the phases such that $i\pi^{x(z)}_{n\uparrow m\downarrow}$ be real.

For the bulk TIs we choose the basis of four states
$\Psi_{v\uparrow}, \Psi_{v\downarrow}, \Psi_{c\uparrow},\Psi_{c\downarrow}$,
where $v$ and $c$ stand for the valence and conduction band, respectively.
This leads to the Hamiltonian \cite{M0_remark}
\begin{eqnarray}\label{Hkp_bulk}
H_{\mathrm{\mathbf{kp}}}&=&C\tau_0\sigma_0+M\tau_z\sigma_0 \\
&-&V_{\|}\tau_x(\sigma_xk_y-\sigma_yk_x)-V_z\tau_y\sigma_0k_z, \nonumber
\end{eqnarray}
with $C=C_0+C_zk_z^2+C_{\|}k_{\|}^2$, $M=M_0+M_zk_z^2+M_{\|}k_{\|}^2$, and
$k_{\|}^2=k_{x}^2+k_{y}^2$.
The 3D TI Hamiltonian (\ref{Hkp_bulk}) is the same (to within a unitary transformation)
as in Refs. \cite{Zhang_Nat_Phys_2009, Liu_PRB_2010, Lu_PRB_2010, Shan_NJP_2010, Zhang_PRB_2012}
(the explicit matrix form is presented by Eq.~(S1) in the SM \cite{SM}).
The Pauli matrix $\bm{\tau}$ operates in valence-conduction band space, whereas
$\bm{\sigma}$ refers to the total angular momentum $\mathrm{\mathbf{J}}$.
In Eq.~(\ref{Hkp_bulk}) a direct product of these matrices is implied.

For the bulk crystals, the parameters in Eq.~(\ref{Hkp_bulk}) are listed in
Table~\ref{tab:table1}, and the eigenvalues $E({\mathbf k})$ of the resulting
four-band Hamiltonian are shown in Figs.~\ref{kp_vs_LDA}(a)--\ref{kp_vs_LDA}(c)
by red lines. Note that already this minimal dimension of $H_{\mathrm{\mathbf{kp}}}$
produces an absolute gap in the spectrum, and for \bise\, and \bite\, its width is
very close to that obtained with the much more accurate eight-band Hamiltonian
[see Eq.~(S3) in the SM and blue lines in Fig.~\ref{kp_vs_LDA}(a)--\ref{kp_vs_LDA}(c)].
An important point about the parameters of the Hamiltonian
(\ref{Hkp_bulk}) is that they are very sensitive to details of the crystal geometry,
as is the {\it ab initio} band structure
\cite{Nechaev_PRBR_2013,Nechaev_PRB_2013,Nechaev_PRB_2015}: even a small variation
in atomic positions leads to considerable changes of the parameters (see Table S1 in the SM \cite{SM}).
Furthermore, in all the 3D systems considered, see Table~\ref{tab:table1}, the
parameters of $H_{\mathrm{\mathbf{kp}}}$ turned out to meet the conditions of the existence of the topological surface states \cite{Shan_NJP_2010}: the diagonal dispersion term $M_{z(\|)}$ is positive and is larger than the
electron-hole asymmetry: $M_{z(\|)}>|C_{z(\|)}|$, although $C_{z}$ and $C_{\|}$ are not
negligible contrary to the assumption in Ref.~\cite{Orlita_PRL_2015}. Thus, our
{\it ab initio} \textbf{k$\cdot$p} Hamiltonian correctly predicts the topological character
of these crystals in accord with the $\mathbb{Z}_2$ topological invariant $\nu_{\rm 3D}$
obtained from the parities of the wave functions at the TRIM points \cite{Fu_PRB_2007}.

\begin{table}
\caption{\label{tab:table1} Parameters of the four-band
\textbf{k$\cdot$p} Hamiltonian (\ref{Hkp_bulk}) for the bulk TIs. We use Rydberg atomic units: $\hbar=2m_0=e^2/2=1$.}
\begin{ruledtabular}
\begin{tabular}{lccccc}
                                 &   \bise    &     \bite    &   \sbte     \\
  \hline
   $V_{\|}$ (a.u.)        &    0.349  &     0.556  &    0.513   \\
   $V_{z}$ (a.u.)        &    0.255  &     0.125  &    0.163   \\
   $C_{0}$ (eV)         &    0.048   &   -0.123  &    0.023   \\
   $C_{z}$ (a.u.)       &     0.37    &     0.70    &   -3.73     \\
   $C_{\|}$ (a.u.)       &     3.65    &   40.54    &   -1.83     \\
   $M_{0}$ (eV)        &    -0.169  &    -0.296  &   -0.182   \\
   $M_{z}$ (a.u.)       &     0.88    &     2.43    &    5.81     \\
   $M_{\|}$ (a.u.)       &     7.71    &   46.55    &  13.47     \\

\end{tabular}
\end{ruledtabular}
\end{table}

We now use our second order perturbation theory to calculate the effective $g$-factors
entering the Zeeman term (see Eq.~(S2) in the SM \cite{SM}) that appears in the
presence of static magnetic field \cite{Winkler, KP_book}:
\begin{align}
g_z^{v(c)} &= \frac{2}{im_0}\left(D_{v(c)\uparrow v(c)\uparrow}^{xy}- D_{v(c)\uparrow v(c)\uparrow}^{yx}\right), \nonumber \\
g_{\|}^{v(c)}&=\frac{2}{im_0}\left(D_{v(c)\uparrow v(c)\downarrow}^{yz}- D_{v(c)\uparrow v(c)\downarrow}^{zy}\right). \nonumber
\end{align}
The most important is that the calculated values are one or even two orders
of magnitude larger than the free electron $g$-factor, $g_0\approx2$. (The values obtained
with the four-band \textbf{k$\cdot$p} method are listed in Table~S1 of the SM \cite{SM}
for all the
3D TIs studied.) This result accords with the recent spin resonance measurements of the
effective $g$-factor in \bise\, \cite{Wolos_PRB_2016}: For the magnetic field parallel
to the $c$ axis, the experimental $g_z$ factors are $27.30\pm0.15$ for electrons and
$29.90\pm0.09$ for holes, while for the field perpendicular to the $c$ axis the $g_{\|}$
factors are $19.48\pm0.07$ and $18.96\pm0.04$ for electrons and holes, respectively. In
order to compare our theoretical effective-mass contributions to the $g$-factors with
the experiment, we should restrict to a two-band Hamiltonian in the basis
$\Psi_{v(c)\uparrow}, \Psi_{v(c)\downarrow}$. For \bise, the 2-band results are in good
qualitative agreement with the experiment: our $g_z$ values are 11.6 (17.7) for electrons
and 19.3 (42.4) for holes. For $g_{\|}$ we get 10.4 (16.1) and 12.1 (16.5) for electrons
and holes, respectively. Here, the values obtained with the LDA-relaxed atomic positions are followed
(in brackets) by those with experimental atomic positions.

In contrast to the bulk TIs, for finite-thickness TI films an ambiguous behavior is observed.
For a 2D system, in the basis
$\Psi^{\mathrm{slab}}_{v\uparrow}, \Psi^{\mathrm{slab}}_{c\downarrow},
\Psi^{\mathrm{slab}}_{c\uparrow},\Psi^{\mathrm{slab}}_{v\downarrow}$
our \textbf{k$\cdot$p} Hamiltonian reads (cf. Refs.~\cite{Bernevig_Science_2006, Zhou_PRL_2008}):
\begin{equation}\label{Hkp_slab}
H^{\mathrm{slab}}_{\mathrm{\mathbf{kp}}}=C\tau_0\sigma_0+M\tau_z\sigma_z-V_{\|}\tau_0(\sigma_xk_y-\sigma_yk_x),
\end{equation}
where $C=C_0+C_{\|}k_{\|}^2$, $M=M_0+M_{\|}k_{\|}^2$, and the operator $\bm{\tau}$
refers now to two decoupled sets of massive Dirac fermions. The last term in Eq.~(\ref{Hkp_slab})
ensures the characteristic \textit{spin-orbital texture} of the TI surface states \cite{Zhang_PRL_2013, Cao_NatPhys_2013}.

\begin{figure}[tbp]
\centering
\includegraphics[angle=0,scale=0.8]{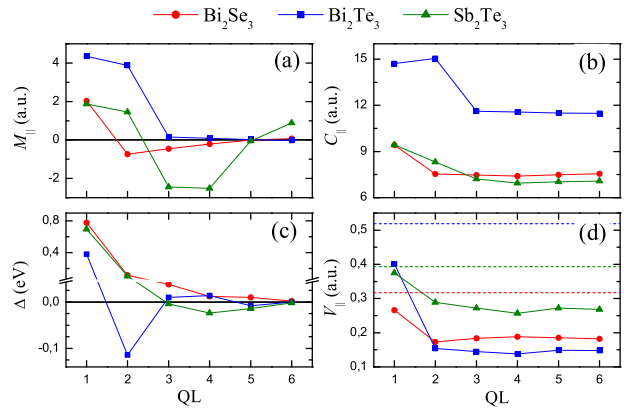}
\caption{The parameters of the Hamiltonian (\ref{Hkp_slab}) for film
thicknesses from 1 QL to 6 QLs. Here, $\Delta=2M_0(-1)^{1+\nu_{2D}}$ is the band gap at
$\bar{\Gamma}$, with $\nu_{2D}$ being the $\mathbb{Z}_2$ topological invariant.
The horizontal dashed lines in the $V_{\|}$ panel show the prediction by the 2D
continuous model in the large thickness limit.} \label{slabs_H_params}
\end{figure}

Figure~\ref{slabs_H_params} shows the film-thickness dependence of the parameters
of the Hamiltonian (\ref{Hkp_slab}) for the TIs considered (the plotted values are
listed in Tables~S2-S4 of the SM \cite{SM}). In contrast to the bulk TIs, for all
the thicknesses the 4-band \textbf{k$\cdot$p} spectrum does not have the absolute
gap, see the red lines in Figs.~\ref{kp_vs_LDA}(d)--\ref{kp_vs_LDA}(f) for 2QL films
and Figs.~S2--S4 of the SM \cite{SM} for other thicknesses.
Note that only for two QLs  and only for \bite\, does the eight-band Hamiltonian
(blue lines) provide a quality close to that achieved in the 3D case.

Velocity $V_{\|}$ and electron-hole asymmetry $C_{\|}$ converge quite fast with
the film thickness [Figs.~\ref{slabs_H_params}(b) and \ref{slabs_H_params}(d)],
demonstrating just slight changes starting from 3~QLs. The 6QL value of $V_{\|}$
can thus be compared with the Fermi velocity from the effective models (see, e.g.,
Ref.~\cite{Lu_PRB_2010}) in the large-thickness limit. In this limit the velocity
is expressed in terms of the \textit{bulk} parameters (Table~\ref{tab:table1})
as $V_{\|}\sqrt{1-C_z^2/M_z^2}$, obviously overestimating the calculated values
[shown by the horizontal dashed line in Fig.~\ref{slabs_H_params}(d)]. The
parameter $C_{\|}$ is positive everywhere,
and it is notably larger than the absolute value of the diagonal dispersion
term $M_{\|}$.

For the same thickness, the parameter $M_{\|}$ may have \textit{different sign} for
different TIs, whereas for a given material $M_{\|}$ is found to ``oscillate'' with
the number of QLs. It turns out that these oscillations do not correlate
with the $\mathbb{Z}_2$ topological invariant $\nu_{\mathrm{2D}}$ obtained from the
parities of the original wave functions at the TRIMs of the 2D BZ. This becomes
evident from a comparison with the behavior of the gap parameter $\Delta$, whose
absolute value $|\Delta|=-2M_0$ yields the gap width at $\bar{\Gamma}$ and sign
depends on $\nu_{\mathrm{2D}}$, with  $\Delta$ being negative for a topologically
non-trivial film.

The \sbte\, film becomes a 2D TI at 3~QLs and preserves this property up
to 6 QLs, Fig.~\ref{slabs_H_params}(c). For \bite, the 2~QL film is non-trivial,
then 3 and 4~QLs are trivial, and 5 and 6~QLs are again non-trivial. The films of
\bise\, are, on the contrary, trivial for all the thicknesses, while the
\textbf{k$\cdot$p} effective models predict them to be a QSHI at some of the
thicknesses, see e.g. Ref.~\cite{Lu_PRB_2010}. It should also be noted that for
the same film the true invariant may depend on details of the crystal geometry
and even on the band structure method, including the choice of
exchange-correlation potential, see
Refs.~\cite{Liu_PRBR_2010, Park_PRL_2010, Kim_PNAS_2012, Bihlmayer_Book_TI, Foerster_PRB_2015, Foerster_PRB_2016, Menshchikova_2016}.
We emphasize that here the topological invariant and the \textbf{k$\cdot$p} parameters
are fully consistent because they are derived from the same  band structure.

\begin{figure}[tbp]
  \centering
  \includegraphics[angle=0,scale=0.75]{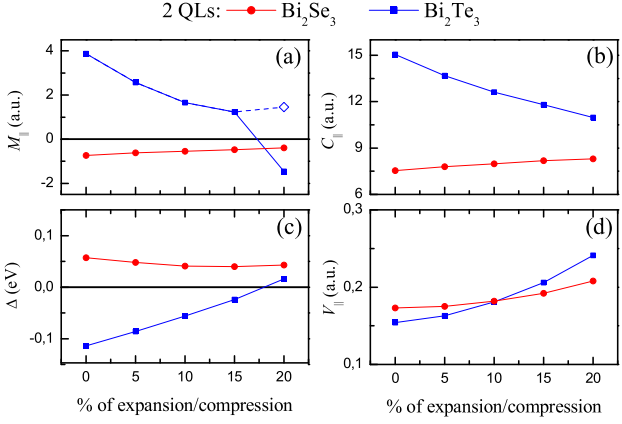}
  \caption{The parameters of the Hamiltonian (\ref{Hkp_slab}) for the 2QL film as a function
of the van-der-Waals spacing.
  The expansion for \bite\, and compression for \bise\, are given in percents of the bulk values.
  The dashed blue line in graph (a) shows the smooth behavior of $\left| M_{\|}\right |$.}
  \label{2QL_H_params}
\end{figure}

According to the effective continuous models \cite{Zhou_PRL_2008, Lu_PRB_2010},
the relation between $C_{\|}$, $M_0$, and $M_{\|}$ one finds in Fig.~\ref{slabs_H_params}
clearly predicts the absence of edge states. This means that a few-band \textbf{k$\cdot$p}
Hamiltonian does not provide a general and certain criterion of the topological character
of 2D systems. Because the electron-hole asymmetry term is sometimes neglected in topological
analysis \cite{Qi_RMP_2011}, it is instructive to consider in more detail the 2QL films of
\bise\, and \bite\, -- the two thinnest films for which the sign of $M_{\|}$ correlates with
the actual $\nu_{\rm 2D}$. In Fig.~\ref{2QL_H_params}, we see that with varying the van-der-Waals
spacing (expansion for \bite\, and compression for \bise) the parameters $V_{\|}$, $C_{\|}$ and
$\Delta$ change steadily, and in \bite\, a transition from QSHI
to the trivial state occurs (at 18\% $\Delta$ becomes positive), and at the same time
$M_{\|}$ becomes negative, again following the true indicator $\nu_{\rm 2D}$.

Finally, let us consider the effective-mass contribution to the $g$-factor for the films.
In our approach, the static magnetic field $\mathrm{\mathbf{B}}$ leads to
the following Zeeman term:
\begin{eqnarray}\label{Hkp_slab_Z}
H^{\rm slab}_{\mathrm{\mathbf{kp}},Z}&=&\frac{\mu_B}{2}\left[g_z\tau_0\sigma_zB_z+g_{\|}\tau_x(\sigma_xB_x+\sigma_yB_y)\right]\\
&+&\frac{\mu_B}{2}\left[\Delta g_z\tau_z\sigma_0B_z+\Delta g_{\|}\tau_y(\sigma_xB_y-\sigma_yB_x)\right], \nonumber
\end{eqnarray}
where $g_{\alpha}=(g_{\alpha}^v+g_{\alpha}^c)/2$ and $\Delta g_{\alpha}=(g_{\alpha}^v-g_{\alpha}^c)/2$.

\begin{figure}[tbp]
\centering
\includegraphics[angle=0,scale=0.75]{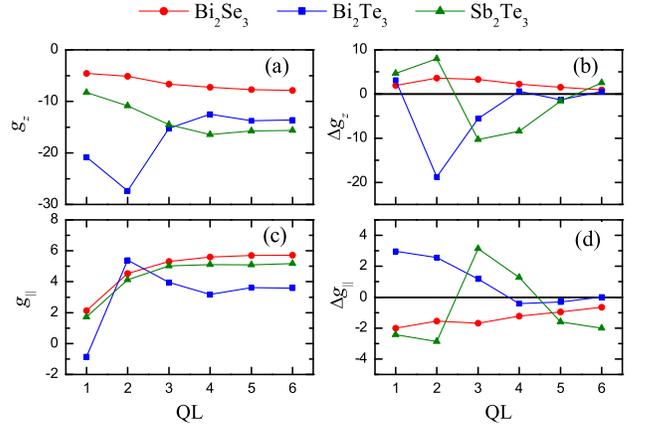}
\caption{The effective Land\'{e} $g$-factors $g_{z(\|)}$ and $\Delta g_{z(\|)}$
entering the Zeeman term (\ref{Hkp_slab_Z}) as functions of the film thickness.}
\label{slabs_Hz_params}
\end{figure}

A novel feature in Eq.~(\ref{Hkp_slab_Z}) is the second term that contains the
$z$-component of the cross product $[\bm{\sigma}\times\mathrm{\mathbf{B}}]$.
It resembles the Zeeman term for inversion-asymmetric quantum wells, where the
``spin-momentum locking'' term is also present \cite{Winkler}. As follows from
Figs.~\ref{slabs_Hz_params}(b) and \ref{slabs_Hz_params}(d), the parameters
$\Delta g_{z}$ and $\Delta g_{\|}$ may ``oscillate'' with the thickness, and for
\bise\, and \bite\, they tend to zero with increasing thickness. As a result, the
leading contribution comes from the ``conventional term'' with $g_{z}$ and $g_{\|}$,
which at 6 QLs is already well converged, Figs.~\ref{slabs_Hz_params}(a) and
\ref{slabs_Hz_params}(c). For \sbte\,, $\Delta g_{z}$ and especially $\Delta g_{\|}$
are rather big at 6 QLs, so the relevant term in $H^{\rm slab}_{\mathrm{\mathbf{kp}},Z}$
should be taken into account at least up to this thickness. Note that in moving from
1 to 6~QLs $\Delta g_{z}$ becomes negative for the first time when the film becomes
a 2D TI, thus demonstrating a correlation with the topological invariant (see also
the effect of expansion/compression for the 2~QL films in Fig. S8 of the SM \cite{SM}).

To summarize, we have developed a fully \textit{ab initio}
\textbf{k$\cdot$p}-perturbation approach to generate model Hamiltonians of a
desired size. This ensures a physically meaningful behavior of the model Hamiltonian
parameters with the continuously varying geometry (van-der-Waals spacing) and for
different number of the building layers. By applying our approach to \bise,
\bite, and \sbte\, films, we have demonstrated that the widely used effective
continuous models are not able to systematically predict the values and often
even the relative sign of the model parameters. The failure to infer the general
and certain criterion from $H_{\mathrm{\mathbf{kp}}}$ stems from its fundamental
limitation: the topological properties of a crystal cannot be unambiguously
determined from the behavior of a few bands in the vicinity of $\mathbf{k}=0$,
even though the band inversion occurs just at that point.

\begin{acknowledgments}
This work was supported by the Spanish Ministry of Economy and
Competitiveness MINECO (Project No. FIS2013-48286-C2-1-P) and Saint
Petersburg State University (Grant No. 15.61.202.2015).
\end{acknowledgments}


\begin{thebibliography}{00}
\bibitem{M0_remark} Note that because the basis functions
  explicitly refer to the valence and conduction band rather than to
  atomic orbitals the parameter $M_0<0$ that defines the band gap does
  not change sign upon moving from the topologically non-trivial
  insulator to the trivial one.
\end{thebibliography}
\end{document}